\documentclass[12pt]{iopart}
\usepackage{epsfig}
\newcommand{\hb} {\mbox{\sffamily HERA \protect\rule[.5ex]{1.ex}{.11ex} B}}
\newcommand{\ra} {\mbox{$\mskip 3mu \rightarrow \mskip 5mu$}}
\newcommand{\mt} {\mbox{$p_\mathrm{T}$}}
\newcommand{\avpt} {\mbox{$< p_\mathrm{T}>$}}
\newcommand{\xeff} {\mbox{$x_\mathrm{F}$}}

\newcommand{\PJgy}{\mbox{\ensuremath{\mathrm{J}\mskip -2mu/\mskip -2mu\psi}}} 
\newcommand{\PJzs}{\mbox{\ensuremath{\mathrm{\psi(2S)}}}}               
\newcommand{\PChi}{\mbox{\ensuremath{\mathrm{\chi_c}}}}                 
\newcommand{\PChin}{\mbox{\ensuremath{\mathrm{\chi_{c0}}}}}             
\newcommand{\PChii}{\mbox{\ensuremath{\mathrm{\chi_{ci}}}}}             
\newcommand{\PUps}{\mbox{\ensuremath{\mathrm{\Upsilon}}}}               
\newcommand{\bbbar}{\mbox{\ensuremath{\mathrm{b\overline{b}}}}}         
\newcommand{\dilepton}{\mbox{\ensuremath{\ell^+ \ell^-}}}               
\newcommand{\epem}{\mbox{\ensuremath{\mathrm{e}^+ \mathrm{e}^-}}}       
\newcommand{\mpmm}{\mbox{\ensuremath{\mu^+ \mu^-}}}                     

\begin{document}

\title{Quarkonia production with the \hb\ experiment}

\author{Joachim Spengler for the \hb\ collaboration\footnote{Talk given at the Quark Matter 2004}}

\address{Max-Planck-Institut f\"ur Kernphysik, Postfach 103989, 69029 Heidelberg, Germany}
\ead{joachim.spengler@desy.de}

\begin{abstract}
Measurements of the dependence of the \PJgy\ production cross section on its
kinematic variables as well as on the target atomic numbers
for 920 GeV/c protons incident on different targets have been made with
the \hb\  detector. The large collected di-lepton sample allows to study the
production ratio of \PJzs\ to \PJgy\  and of \PChi\  to \PJgy . 
We also report on measurements of the \bbbar\ and \PUps\ production cross
section.
\end{abstract}

\submitto{\JPG}

\section{Introduction}

\hb\ is a fixed-target experiment which studies collisions of
protons with the nuclei of atoms in target wires positioned in the
halo of HERA's 920 GeV proton beam. The large acceptance of the \hb\ 
spectrometer coupled with high-granularity particle-identification
devices and a precision vertex detector allow for detailed studies of
complex multi-particle final states. \hb 's sophisticated di-lepton
trigger is capable of finding rare interactions containing two leptons
coming, for example, from the decay of a \PJgy\ amidst an
overwhelming background of more ordinary proton-nucleus interactions.
By using target wires made from a variety of materials (carbon,
aluminum, titanium and tungsten) \hb\ is able to study the
dependence of various properties of proton-nucleus interactions as a
function of atomic number.

Analysis of modest data samples which had been taken in the
commissioning run of year 2000 was completed by early 2003 and led to
publications on the \bbbar\ cross section~\cite{herab1}, the ratio of
\PChi\ to \PJgy\ production cross sections~\cite{herab2}  and on the
production cross sections of $\Lambda^0$ and $\mathrm{K}^0_s$~\cite{herab3}.

During the HERA run which began in Summer 2002 and ended in March
2003, \hb\ accumulated 150 million events with its di-lepton
trigger.  The di-lepton sample includes 300,000 events with a
\PJgy\ in the final state, roughly equally divided between
di-muon and di-electron decays. This sample is more than 50 times
larger than the sample accumulated in the 2000 run, but thanks to the
large number of improvements in detector and trigger systems could be
recorded in a similar amount of beam time.  Thanks to improvements in
the data acquisition system, recording rates of a 1000 events per
second (more than 1 Terabyte per day) for minimum bias data were
achieved which allowed collection of a 200~million events minimum bias
sample in just one week of calendar time.  This represents an increase
in sample size by more than a factor of 100 with respect to the 2000
run.  Having successfully accumulated these interesting data samples,
the \hb\ Collaboration decided in early 2003 to definitively finish
data-taking activities at the end of running period and to concentrate
on analyzing the accumulated data.  Several studies are in progress on
both the di-lepton and the minimum bias samples.

\section{Preliminary results}

\subsection{The Di-lepton sample}

\begin{figure}
\begin{center}
\epsfig{file=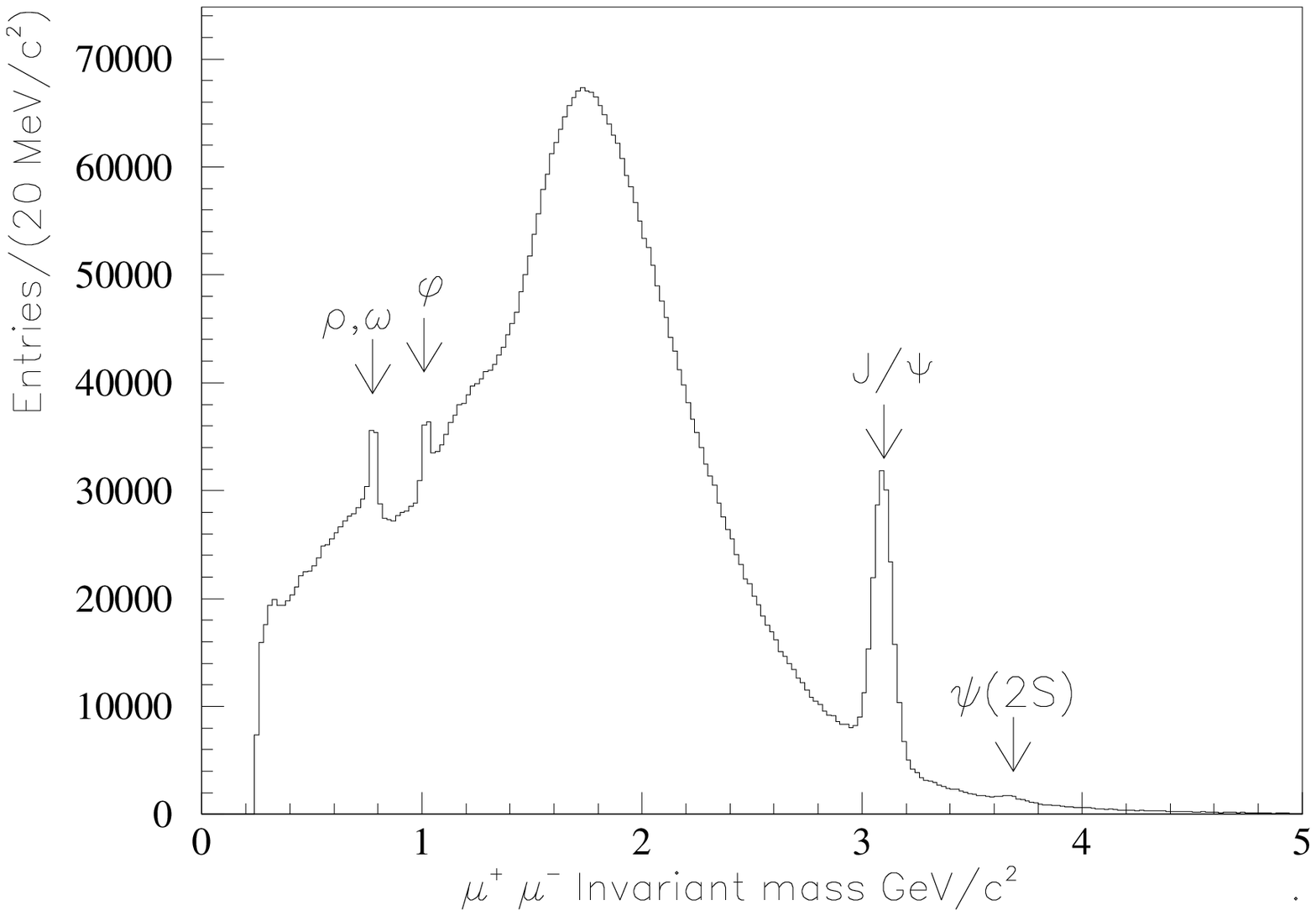,height=5.0cm}
\epsfig{file=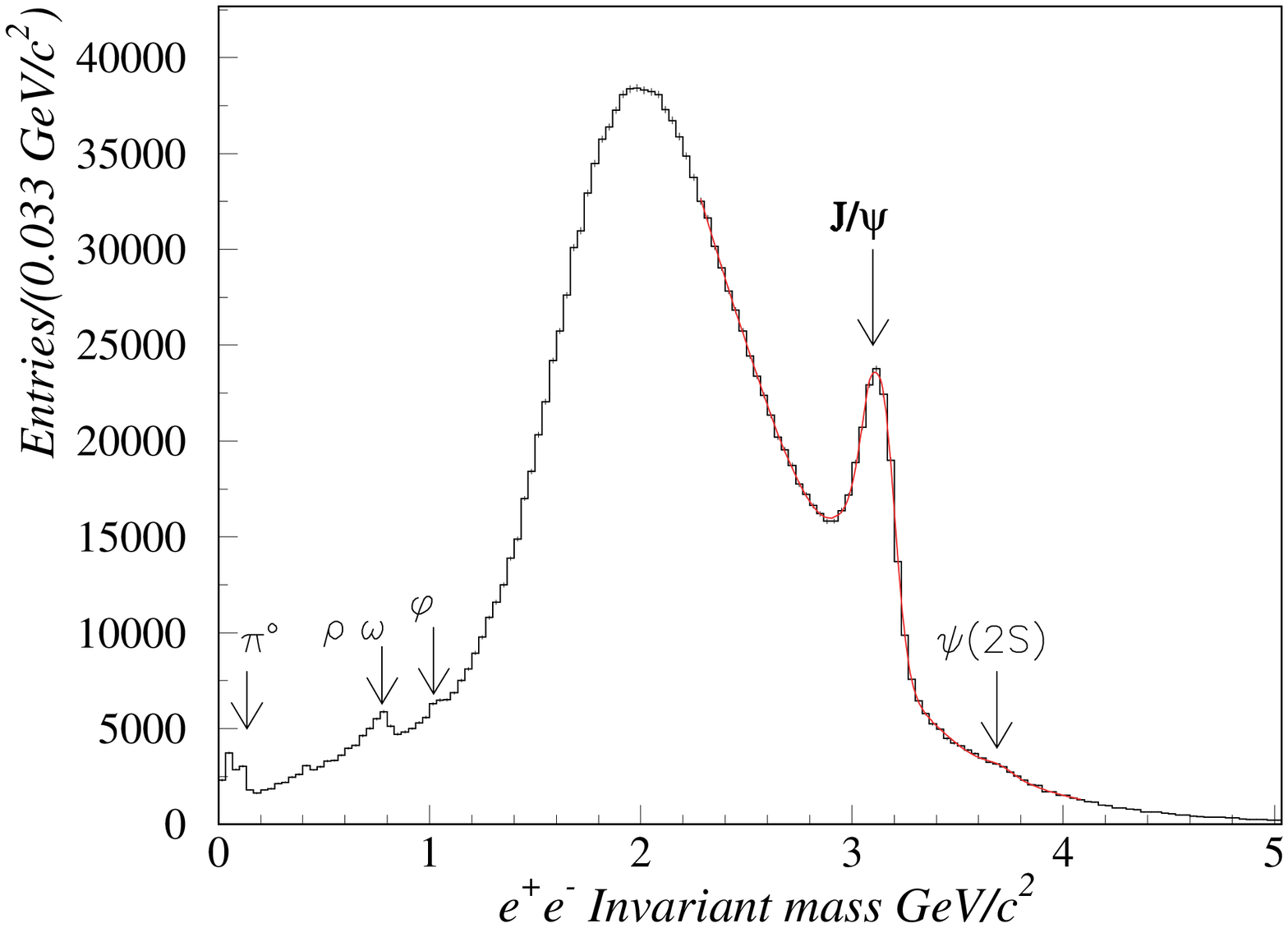,height=5.2cm,bbllx=30,bblly=233,bburx=587,bbury=638}
\end{center}
\caption{ The \mpmm\ (left) and \epem\ (right)
invariant mass spectrum.}
\label{fig:jpsimumu}
\end{figure}

The \mpmm\ invariant mass plot from 80\% of the full
sample is shown in Fig.~\ref{fig:jpsimumu}. A prominent peak containing
150,000 di-muon pairs at the mass of the \PJgy\ is
seen. Other features of the plot include peaks at the masses of the
\PJzs\ , the $\phi(1020)$, the $\omega(782)$ and the
$\rho(770)$. The same features, although at lower resolution due to the
bremsstrahlung emitted by the electrons, can
be seen in the \epem\ mass plot shown in Fig.~\ref{fig:jpsimumu}.
The \PJgy\ \ra\ \epem\ peak contains 100,000 events.

\subsection{Differential distributions of \PJgy\ }

\begin{figure}
\begin{center}
\begin{minipage}{0.45\textwidth}
\begin{center}
\epsfig{file=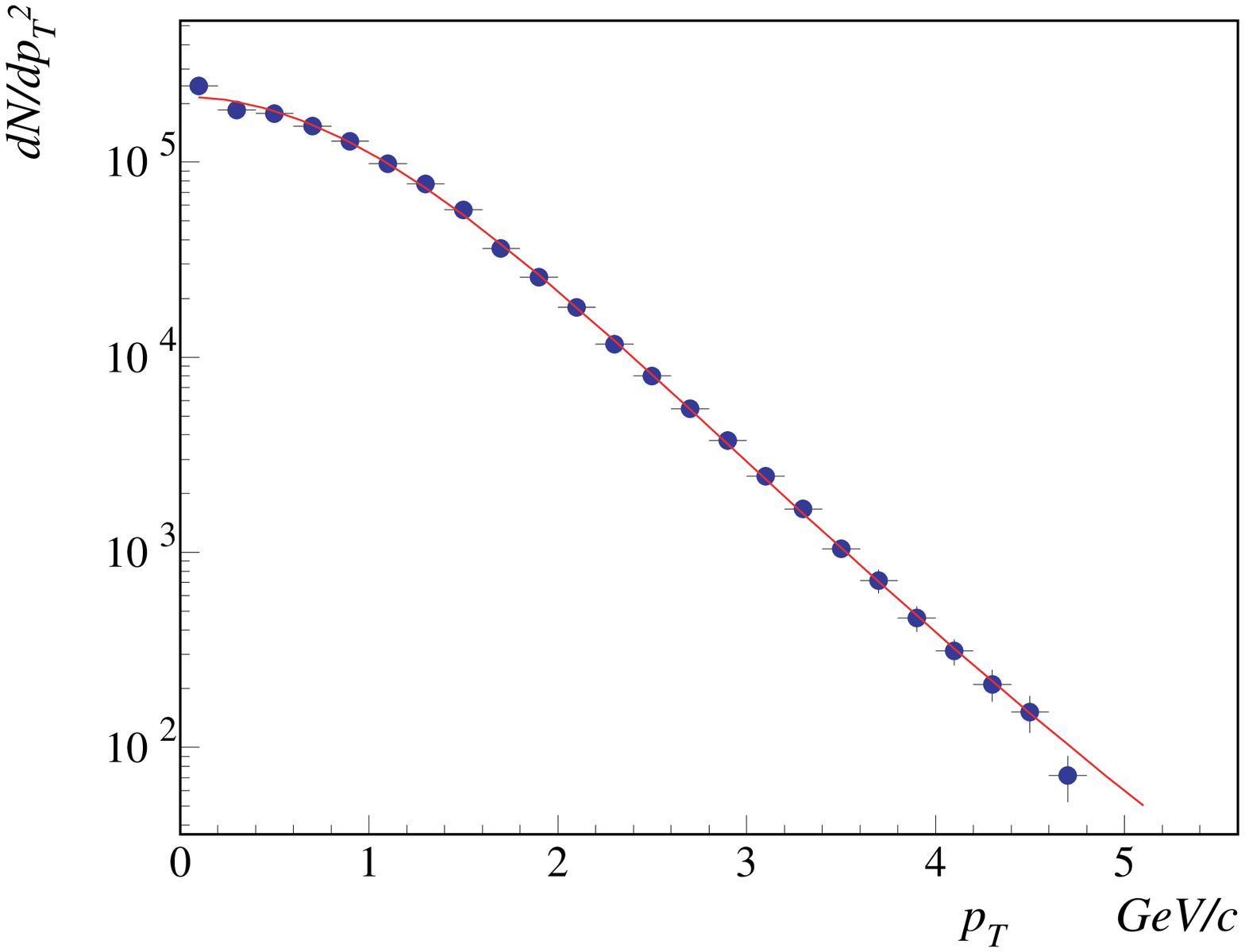,height=5.5cm}
\end{center}
\caption{The \mt\ distribution of \PJgy\ \ra\ \epem\  .}
\label{fig:ptjpsi}
\end{minipage}
\begin{minipage}{0.45\textwidth}
\begin{center}
\hspace*{5mm}
\epsfig{file=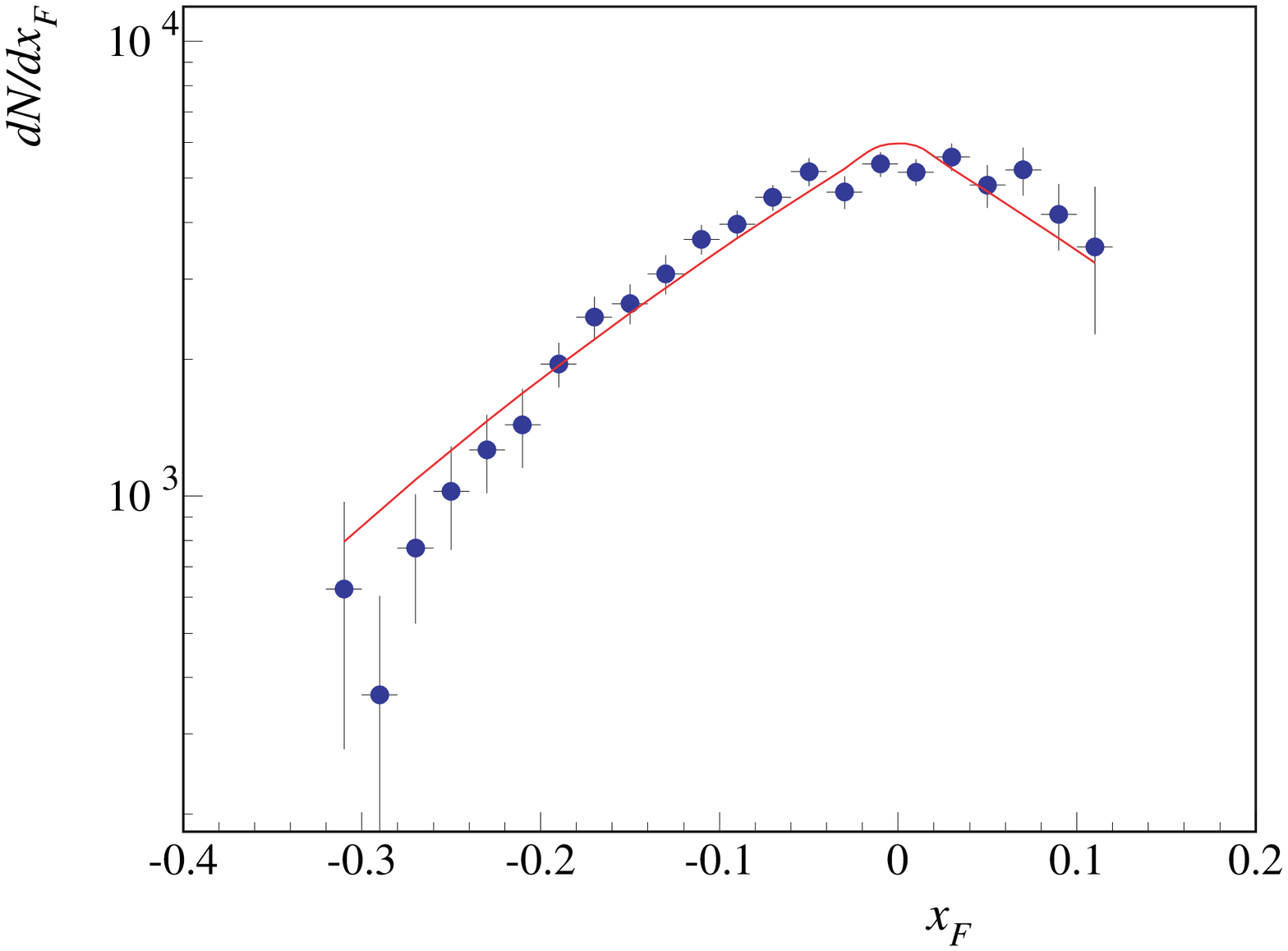,height=5.45cm}
\end{center}
\caption{The \xeff\ distribution of \PJgy\ \ra\ \epem\  .}
\label{fig:xfjpsi}
\end{minipage}
\end{center}
\end{figure}

The large data sample in connection with the large acceptance of
the \hb\  spectrometer allows for a precise determination of the
kinematical distributions of \PJgy . The \mt -distribution is
plotted Fig.~\ref{fig:ptjpsi} for 80\% of the \epem  sample. Note
the wide range of up to 4.8 GeV/c in transverse momentum.

The average \mt\  is determined by fitting the following expression to the data
\begin{displaymath}
   \frac{d \sigma}{d \mt ^2  } =  \beta \cdot
   \left[ 1 + \left( \frac{35 \cdot \pi \cdot \mt }{256 \; \cdot < \mt > } \right)^2
   \right] ^{-6} . 
\end{displaymath}
The outcome of this analysis can be found in Table~\ref{tab:tptjpsi}. The \hb\
data confirm the well known trend of \mt -broadening with increasing
atomic number as it was reported during this conference for
example by Phenix~\cite{Phenix}.

\begin{table}[ht]
\caption{\label{tab:tptjpsi}Average transverse momentum of \PJgy . For the
\hb\ results only statistical errors are given.}
\begin{indented}
\item[]\begin{tabular}{@{}cccccc}
\br
   Target & Exp.  & Range in \mt & \avpt ($\mathrm{e}^{\pm}$) & \avpt($\mu^{\pm}$) & error \\ 
\mr
C 920  & Hera-B & $\leq$ 4.8 & 1.22 & 1.22 & 0.01 \\
W 920  & Hera-B & $\leq$ 4.8 & 1.29 & 1.29 & 0.01 \\
Si 800 & E771~\cite{E771a} & $\leq$ 3.5 &      & 1.20 & 0.01 \\
Au 800 & E789~\cite{E789a} & $\leq$ 2.6 &      & 1.29 & 0.009 \\
\br
\end{tabular}
\end{indented}
\end{table}

The range in \xeff\  of \PJgy\  which is accessible with the \hb\  detector
can be seen from  Fig.~\ref{fig:xfjpsi}. It is for the first time
that \PJgy -production
can be measured in the backward region.
However, due to still ongoing acceptance
calculations, only a small fraction of about 10\% of the \epem\ sample
is shown in this plot. In order to compare this distribution with
other results, the expression
\begin{displaymath}
   \frac{d \sigma}{d \xeff  } =  \beta \cdot ( 1 - | \xeff | )^c 
\end{displaymath}
was fitted to our data. For the time being, only a range for the slope c can
be quoted (see Table~\ref{tab:txfjpsi}).

\begin{table}[ht]
\caption{\label{tab:txfjpsi}Exponent c of \xeff\  distribution of \PJgy .
For the \hb\  results only statistical errors are given.}
\begin{indented}
\item[]\begin{tabular}{@{}cccccc}
\br
 Target & Exp.  & Range in \xeff  & c($\mathrm{e}^{\pm}$) & c($\mu^{\pm}$)  & error \\ 
\mr
C, W 920  & Hera-B & -0.35 $\leq$ \xeff  $\leq$ 0.15 & 5 - 6.5 &  & 0.3 \\
Si 800 & E771~\cite{E771a} &-0.05 $\leq$ \xeff  $\leq$ 0.25  &      & 6.54 & 0.23 \\
Au 800 & E789~\cite{E789a} &-0.03 $\leq$ \xeff  $\leq$ 0.13  &      & 4.91 & 0.18 \\
Cu 800 & E789~\cite{E789b} &0.30 $\leq$ \xeff  $\leq$ 0.95   &      & 5.21 & 0.04 \\
\br
\end{tabular}
\end{indented}
\end{table}

\begin{figure}[ht]
\begin{center}
\epsfig{file=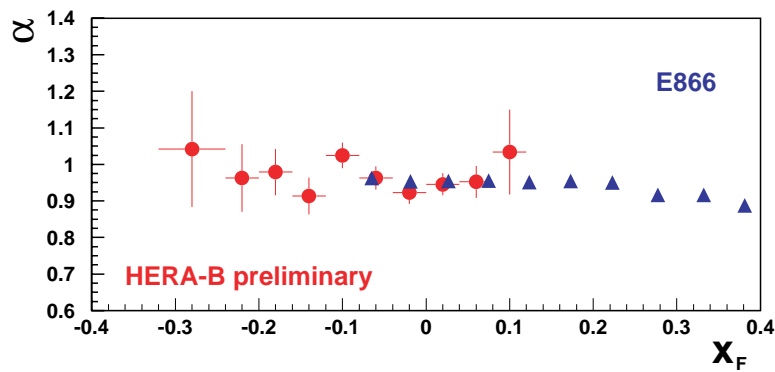,height=5.0cm}
\end{center}
\caption{$\alpha$ versus \xeff\  for \PJgy . The \hb\  points were normalized
to the E866 data in the region of overlap. }
\label{fig:xfadep}
\end{figure}

Due to the simultaneously recorded data sets with carbon and tungsten target,
\hb\ is well suited to study the A-dependence of \PJgy -production, which
is usually parametrized as 
\begin{displaymath}
   \sigma_{pA} =  \sigma_{pN} \cdot  A^{\alpha} ; \; \;
\alpha = \frac{1}{log(A_W/A_C)} \cdot log \left( \frac{N_W}{N_C}
\frac{\mathcal{L}_C}{\mathcal{L}_W}
\frac{\epsilon_C}{\epsilon_W} \right) .     
\end{displaymath}
Thus the measurement of $\alpha$ requires the ratio of the number of
observed \PJgy\ 's, the ratio of luminosities ($\mathcal{L}_{C,W}$)
and the ratio of the reconstruction efficiencies ($\epsilon_{C,W}$)
for the two different target materials. Up to now, the absolute
ratio of luminosities is still under evaluation. Therefore our data
points, determined from 25\% of the \epem\ sample (see Fig.~\ref{fig:xfadep}),
were normalized to
the E866 result~\cite{E866} in the overlap region of the \xeff\  distribution.

\subsection{Production ratio of \PJzs\ to \PJgy\ }

\begin{figure}[t]
\begin{minipage}{0.45\textwidth}
\begin{center}
\epsfig{file=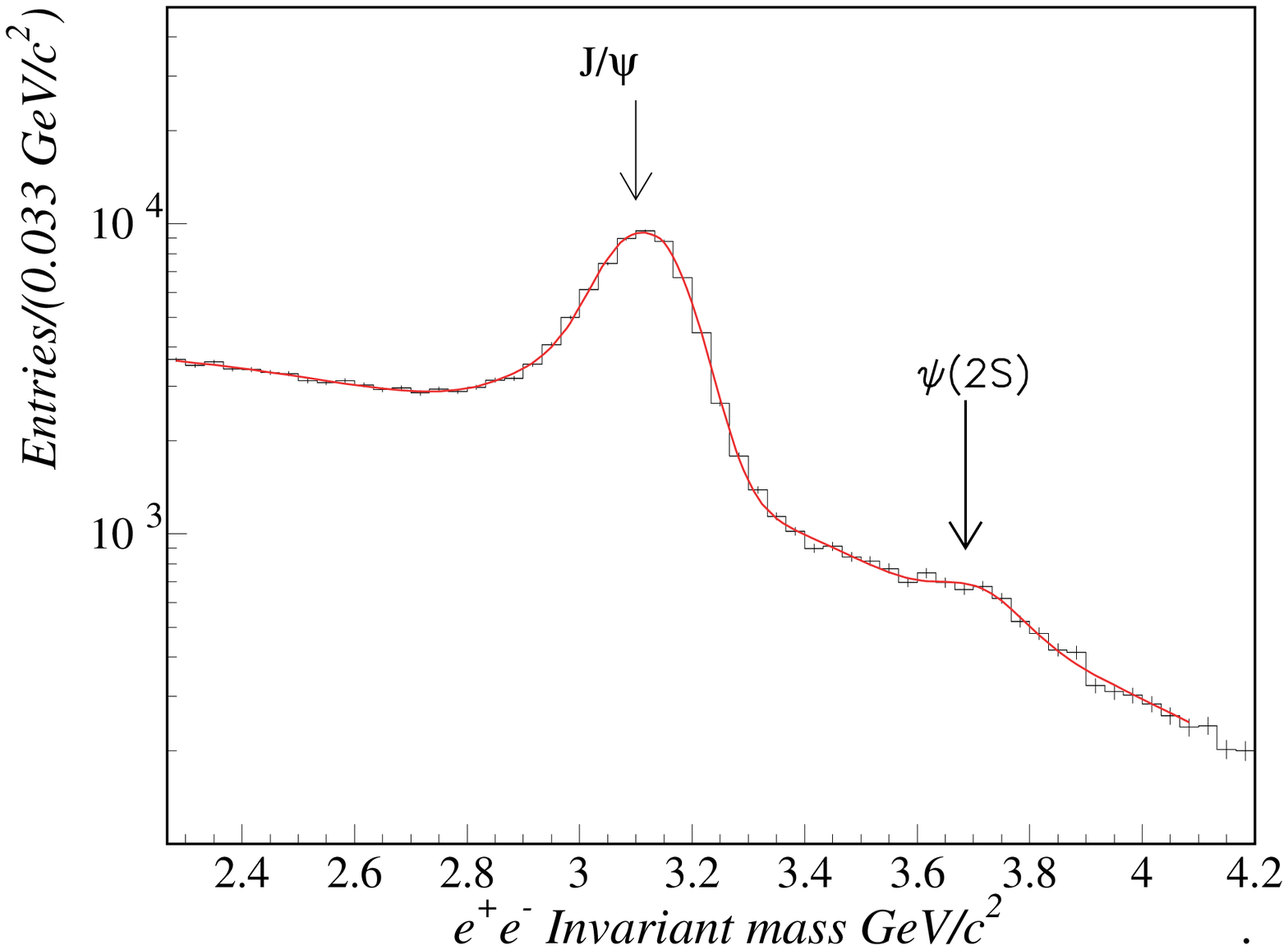,height=5.5cm}
\end{center}
\caption{The \epem\ invariant mass spectrum.}
\label{fig:jpsi(2s)ee}
\end{minipage}
\begin{minipage}{0.45\textwidth}
\begin{center}
\hspace*{5mm}
\epsfig{file=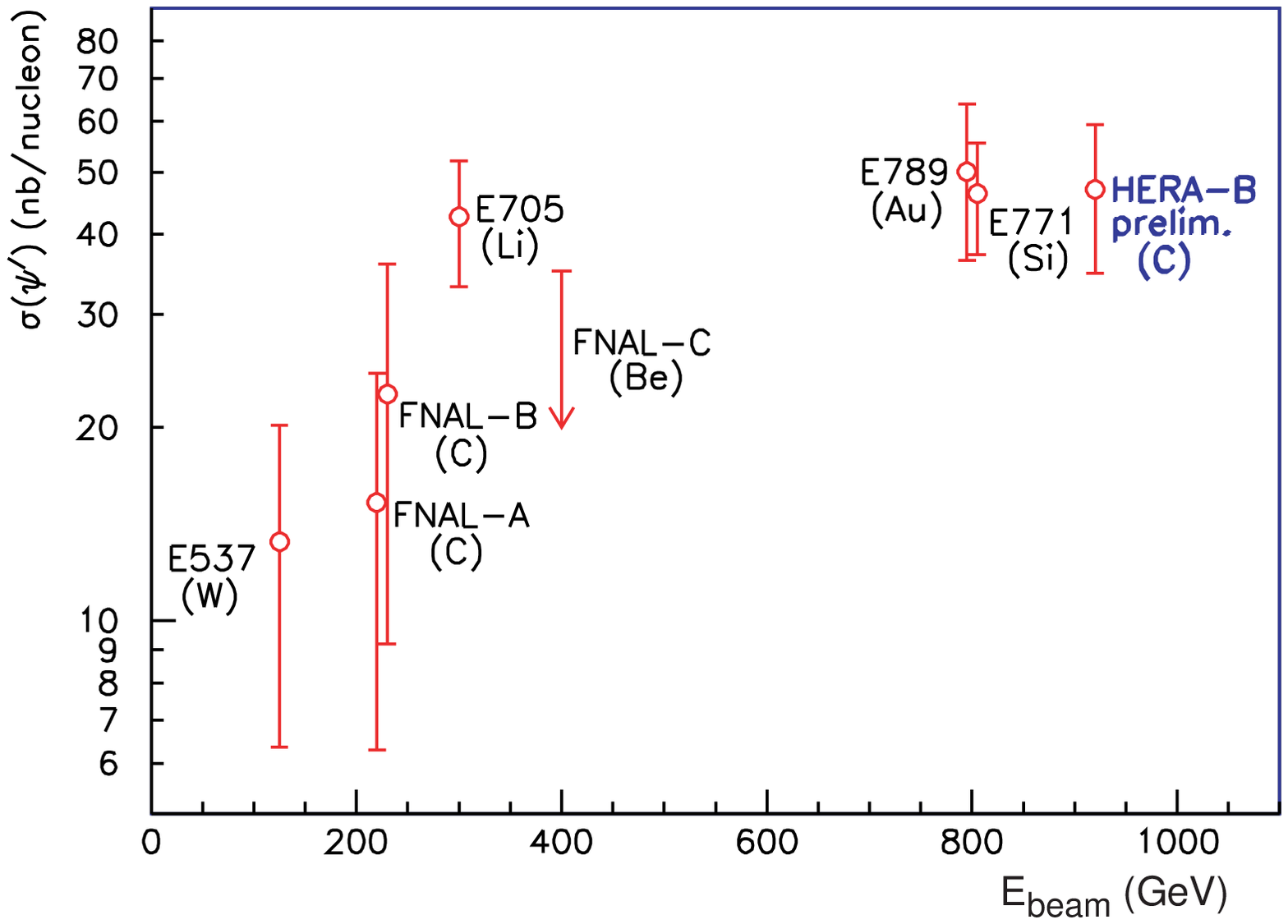,height=6.0cm,bbllx=72,bblly=256,bburx=526,bbury=608}
\end{center}
\caption{The \PJzs\ cross section as function of the proton beam energy.}
\label{fig:expsit2s}
\end{minipage}
\end{figure}

A precise measurement of the \PJzs\ cross section allows to test
the importance of color octet production channels~\cite{Theo6}.
Analyzing 30\% of the \epem\  data sample (Carbon target), a clear \PJzs\ peak
can be seen in  Fig.~\ref{fig:jpsi(2s)ee}. In order to minimize systematic
uncertainties, the following ratio
\begin{displaymath}
   R_{\PJzs\ } = \frac{\sigma(\PJzs\ )}{\sigma(\PJgy\ )} =
   \frac{n(\PJzs\ )}{n(\PJgy\ )} \cdot
   \frac{Br(\PJgy\ \ra\ \epem\ )}{Br(\PJzs\ \ra\ \epem\ )} \cdot   
   \frac{\epsilon(\PJgy\ )}{\epsilon(\PJzs\ )}
\end{displaymath}
is determined. We obtain a result of  $ R_{ \PJzs\ } = 0.13 \pm 0.02 $ which
is in good agreement with the result of the Muon channel. For the
time being, no systematic error can be quoted because studies are ongoing.
Based on results of E771~\cite{bbbar2} and E789~\cite{bbbar1},
we use $R_{\PJzs\ }=(357 \pm 8 \pm 27)$ nb/nucleon~\cite{herab1} to determine
the \PJzs\ production cross section $\sigma(\PJzs ) = (46 \pm 12) $ nb/nucleon.
This preliminary value is shown in Fig.~\ref{fig:expsit2s} together with
results of other experiments.

\subsection{Production ratio of \PChi\ to \PJgy\ }

\begin{figure}[ht]
\begin{center}
\epsfig{file=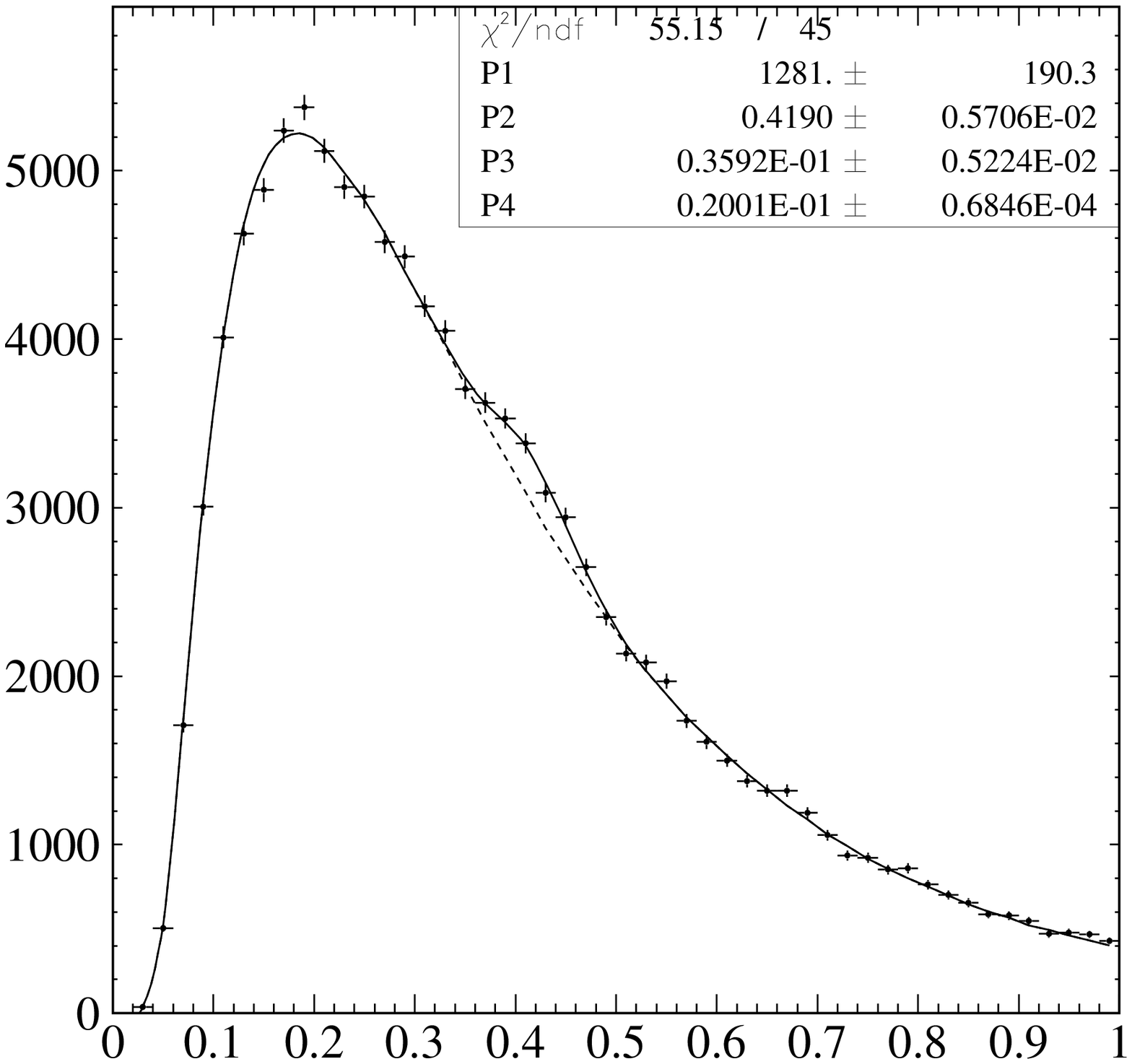,height=5.79cm}
\hspace*{5mm}
\epsfig{file=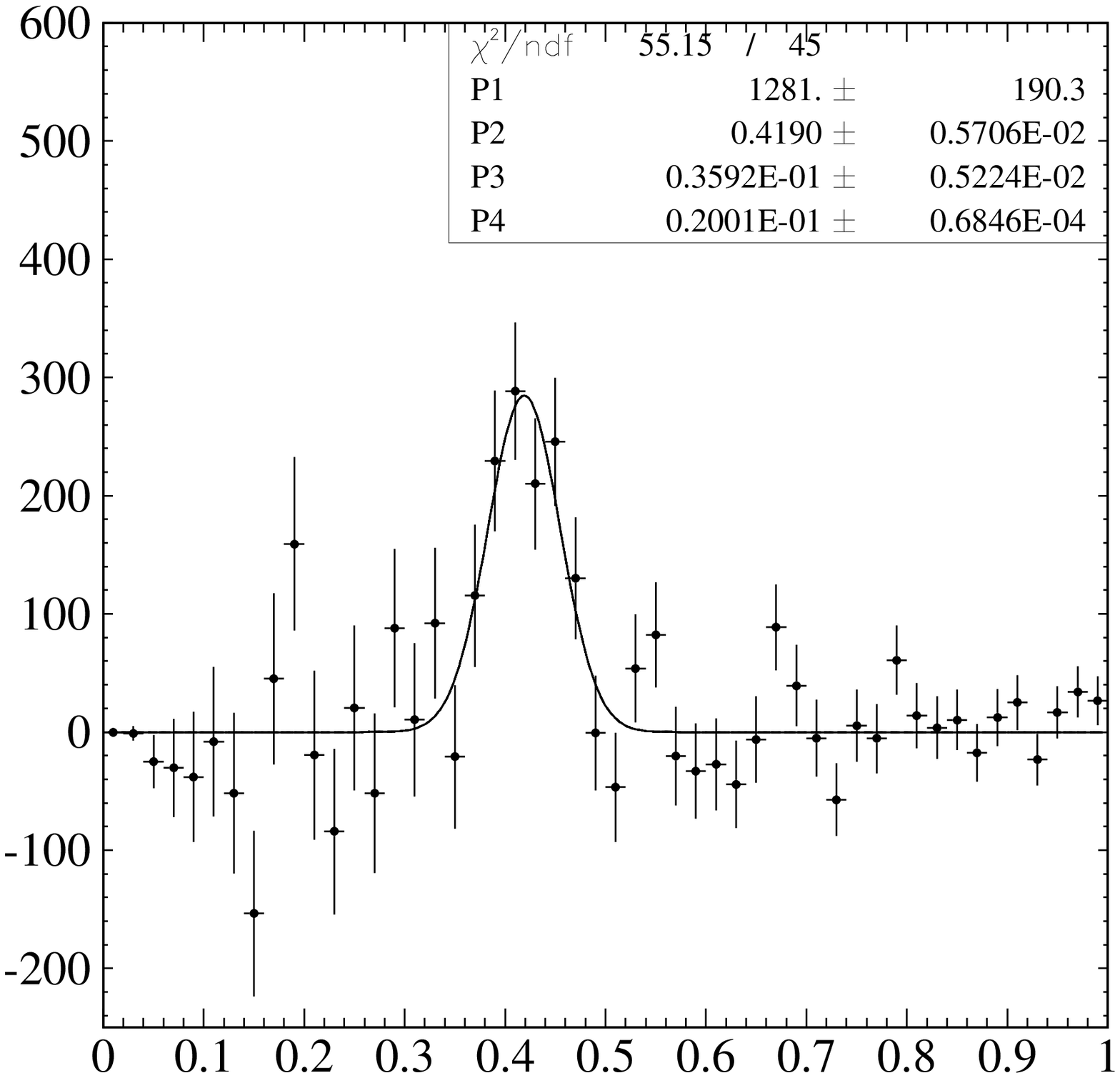,height=6.0cm}
\end{center}
\caption{Spectrum of the mass difference 
$\mathrm{\Delta m = m(l^+l^- \gamma) - m(l^+l^- )}$ before (left)
and after subtraction (right) of the combinatorial background.}
\label{fig:chic}
\end{figure}

A measurement of the production ratio of \PChi\ to \PJgy\ 
will help to choose between models~\cite{Theo1, Theo2, Theo3, Theo7}
attempting to describe the production of quarkonium states.
Using the year-2000 data set, \hb\ measured the fraction of \PJgy\
coming from \PChi\ decays
\begin{displaymath}
  R_{ \PChi\ } = \frac{ \sum_{i=1}^{2} \sigma(\PChii\ ) \cdot Br(\PChii\ 
  \ra\ \PJgy\ \gamma) }{\sigma(\PJgy\ )} =
  \frac{n(\PChi\ )}{n(\PJgy\ )} \cdot 
  \frac{\epsilon (\PJgy\ )}{\epsilon (\PChi ) \cdot \epsilon (\gamma ) }
\end{displaymath}
based on $380 \pm 74$ reconstructed \PChi\  mesons to be 
$ R_{\PChi } = 0.32 \pm 0.06 \pm 0.04 $~\cite{herab2} which is in good
agreement with the NRQCD calculations~\cite{Theo3}.  The state
\PChin\  was neglected due to its small branching ratio.
Fig.~\ref{fig:chic} shows the mass difference 
$\mathrm{\Delta m = m(l^+l^- \gamma) - m(l^+l^- )}$ for 15\% of the 
\mpmm\  data sample recorded during the 2002/03 period.
For the full sample over 10,000 reconstructed \PChi\ mesons are expected
in both final states which will lead to a much more precise measurement.

\subsection{\bbbar\ Cross Section}

A precise measurement of the b cross section in
hadronic interactions at fixed target energies would be welcome in
part as a test of calculations, in part to give guidance for
setting the parameters of the theory \cite{Theo4, Theo5}.

\begin{figure}
\begin{center}
\epsfig{file=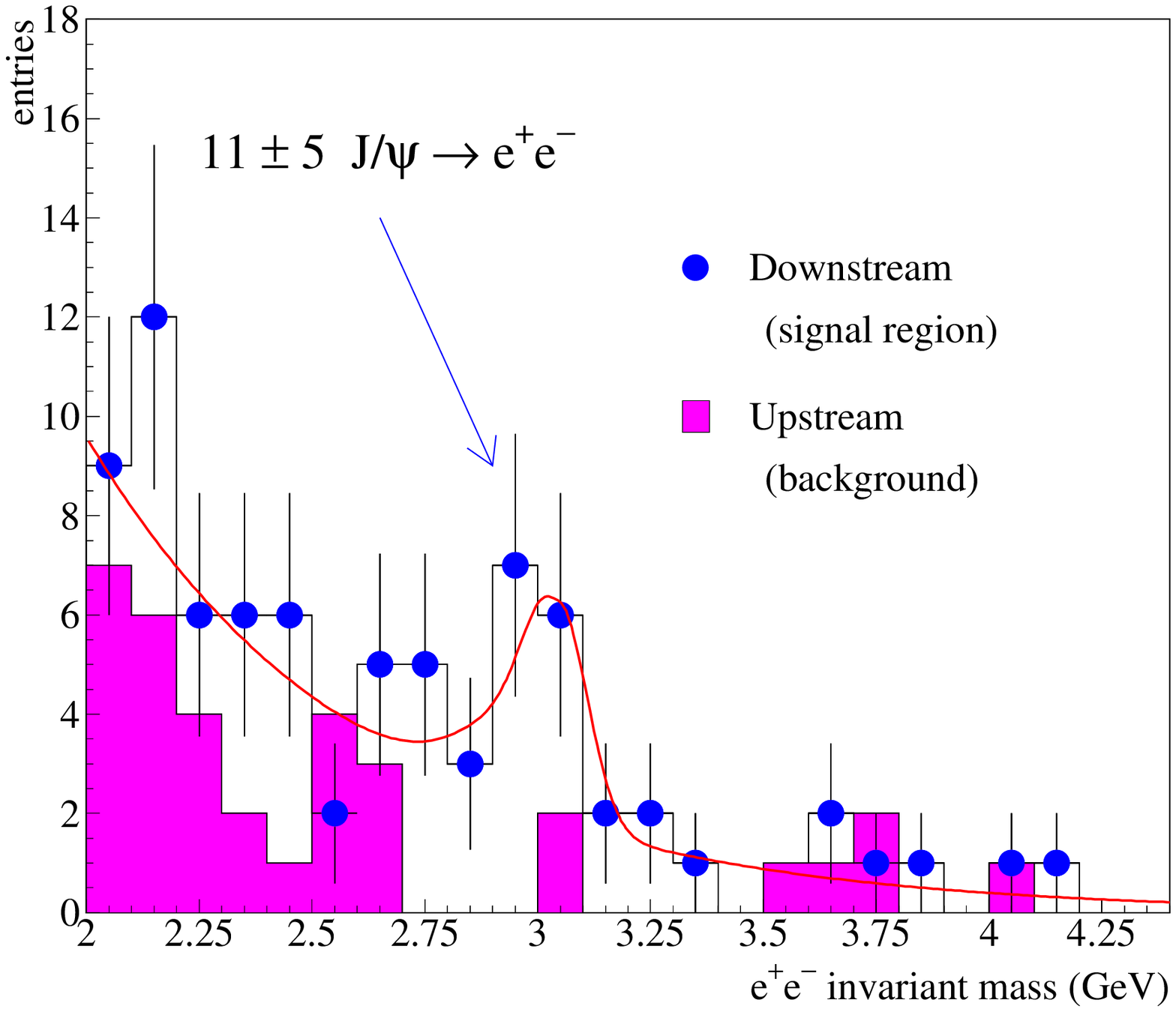,height=6.0cm}
\hspace*{3mm}
\epsfig{file=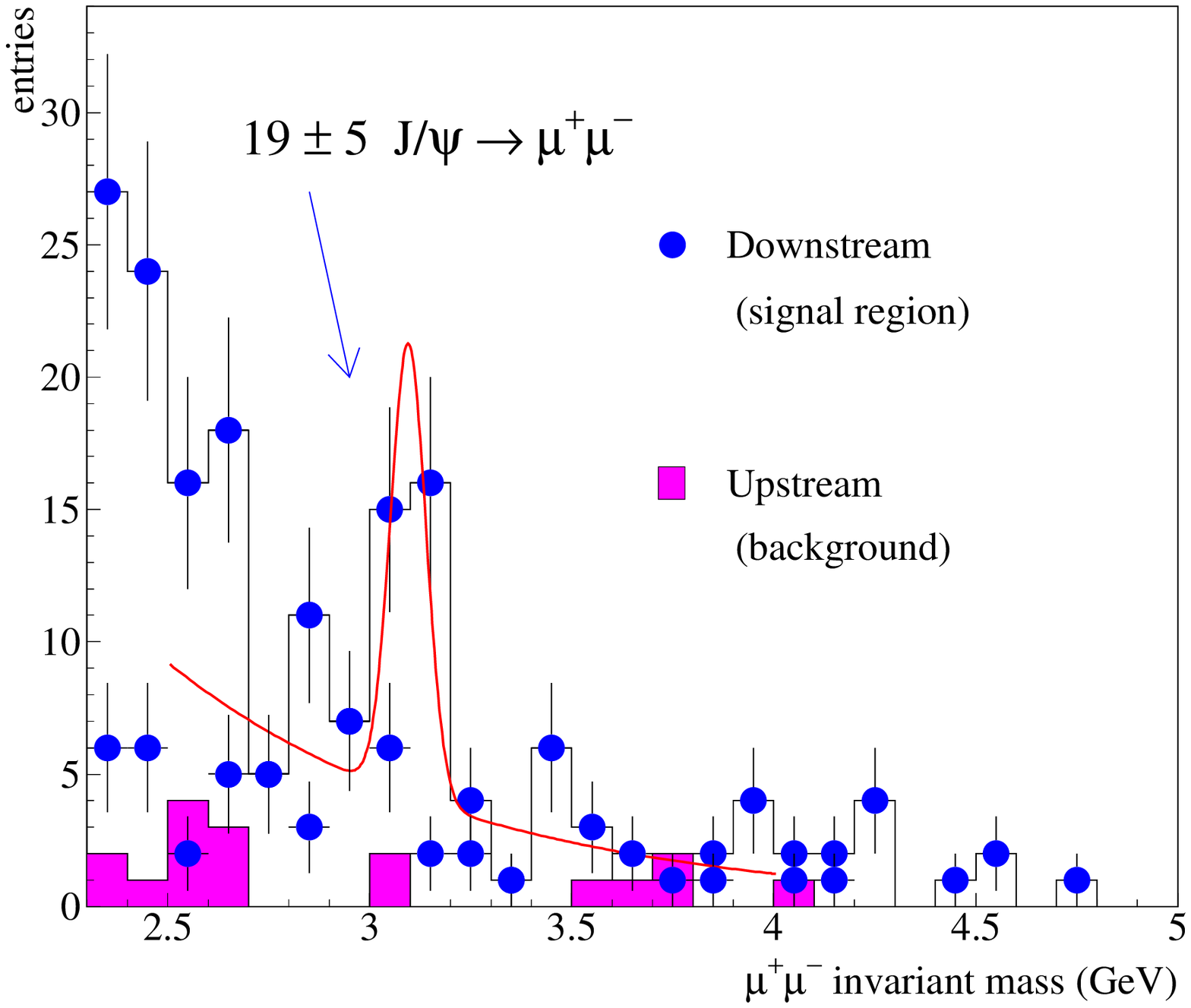,height=6.0cm}
\caption{Invariant mass distribution of \epem  (left) and
\mpmm -pairs (right) after application of vertex cuts. Points with
error bars: lepton pairs downstream of the vertex.
Shaded histogram: lepton pairs which, due to finite measurement resolution,
appear to originate upstream of the main interaction point.}
\label{fig:bbbar}
\end{center}
\end{figure}

\begin{figure}[ht]
\begin{center}
\begin{minipage}{0.45\textwidth}
\begin{center}
\hspace*{-2mm}
\epsfig{file=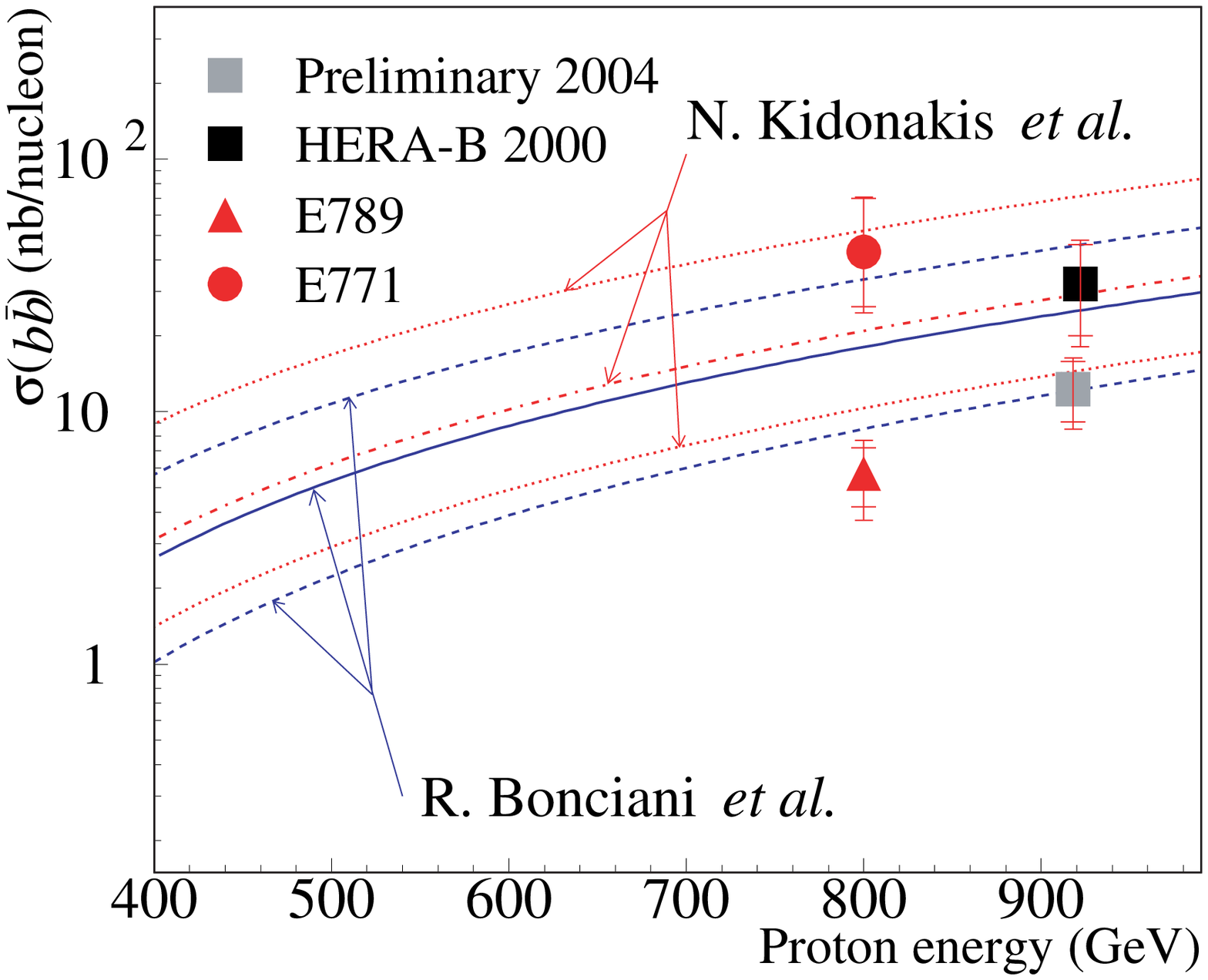,height=6.4cm}
\end{center}
\caption{Comparison of the \hb\ results of 2000 and 2004 
with other experiments and theoretical predictions.}
\label{fig:bpredict}
\end{minipage}
\begin{minipage}{0.45\textwidth}
\begin{center}
\vspace*{3.5mm}
\hspace*{5mm}
\epsfig{file=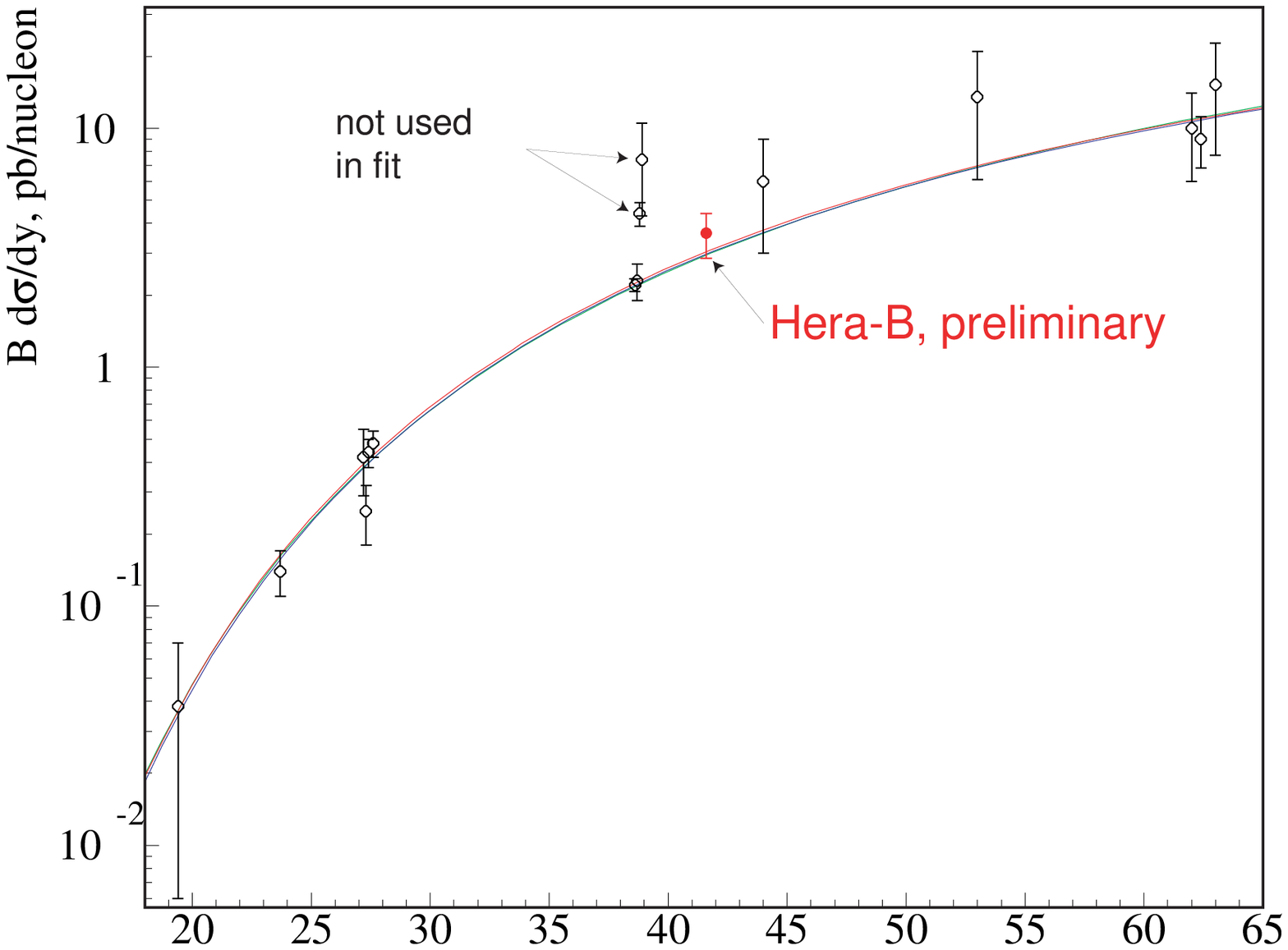,height=5.7cm}
\end{center}
\caption{$Br \cdot \frac{d\sigma}{dy} |_{y=0} $ for \PUps\ production as function
of the center-of-mass energy. The fit to the data is described in the text.} 
\label{fig:Upsfit}
\end{minipage}
\end{center}
\end{figure}

The principle of the \hb\ B cross section
determination is to measure the rate of \PJgy 's which
originate typically 1~cm downstream of the target and compare it to
the rate of directly produced \PJgy 's. Using this technique and
the data collected during the
year-2000 running period, a total \bbbar\  production cross section of
$\sigma (\bbbar ) = (32 \pm 13 \pm 6) $ nb/nucleon
was derived. The much larger new sample will allow a considerably more
precise measurement. Fig.~\ref{fig:bbbar}  shows the \epem\ (left) and
\mpmm\ (right) invariant mass spectra from 35\% of the data sample
after applying vertex cuts to suppress the directly-produced \PJgy 's.  
The downstream samples clearly show the
\PJgy\ but not the upstream samples. For the full sample we
expect approximately 100 \PJgy 's from b decay. Cut
optimization and systematic studies are underway. A preliminary estimate
of the cross section/nucleon, which is $1.5 \sigma$
below the 2000 result, is shown in Fig.~\ref{fig:bpredict} together
with the results of E771~\cite{bbbar2} and E789~\cite{bbbar1}. The
latest QCD calculations~\cite{Theo4, Theo5} are indicated by their central
values as well as upper and lower bounds. 

\subsection{Production cross section of \PUps\ }

\begin{figure}[ht]
\begin{center}
\epsfig{file=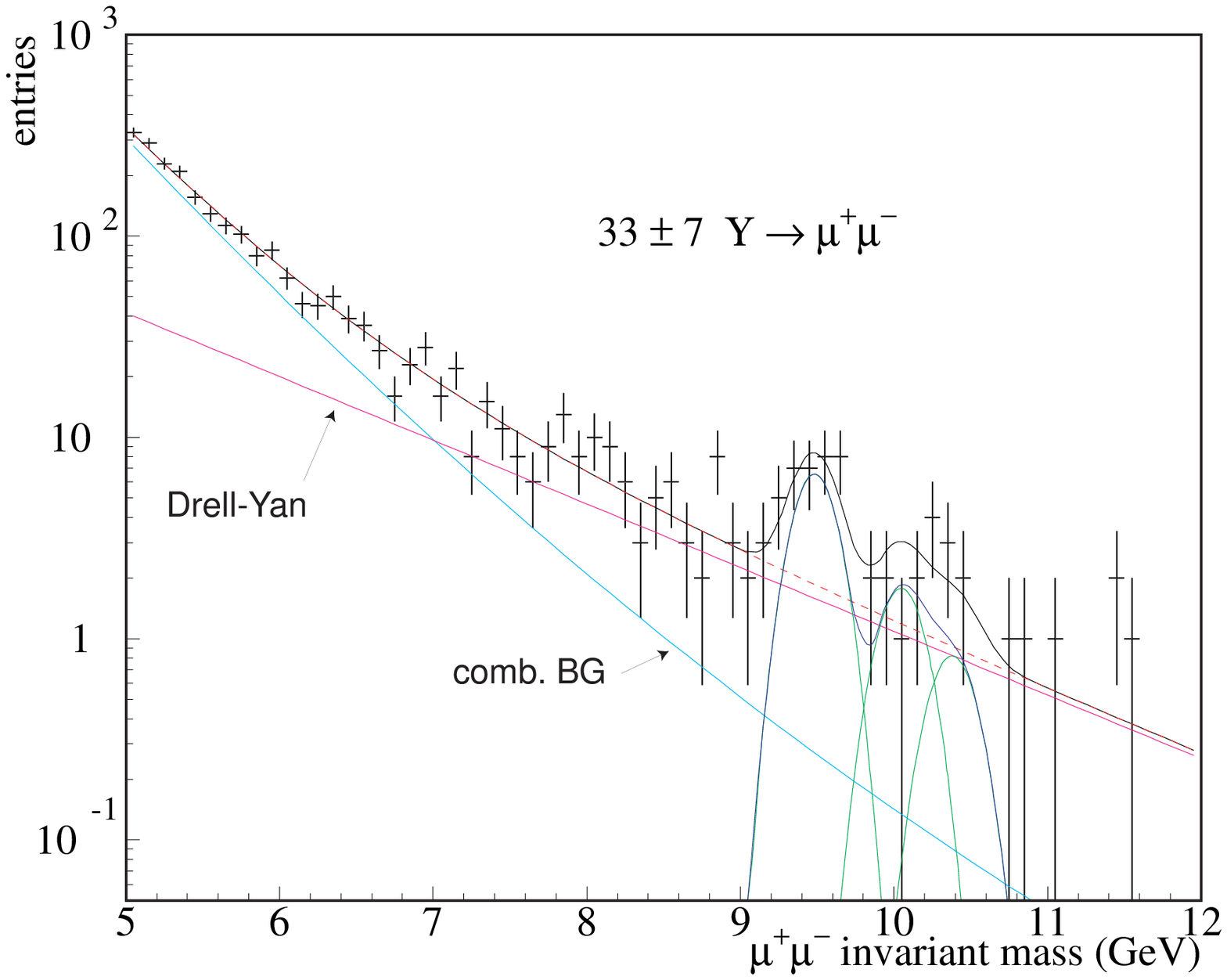,height=6.0cm}
\epsfig{file=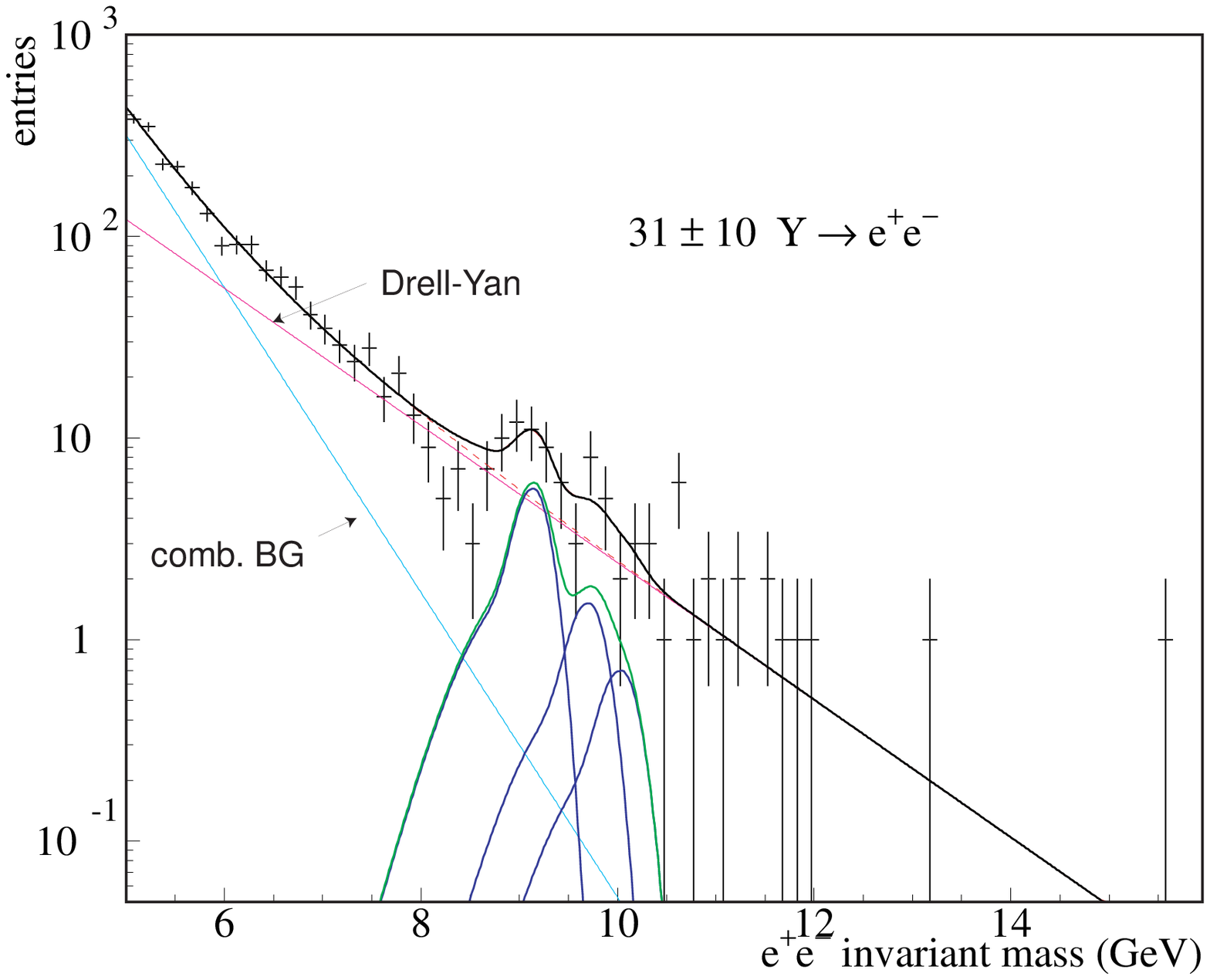,height=6.0cm}
\end{center}
\caption{The \mpmm\ (left) and \epem\ (right) invariant mass spectrum.} 
\label{fig:Upsmmee}
\end{figure}

In order to search for \PUps -mesons, both C and W data samples were combined
assuming an A-dependence of $\alpha = 1$.  
As can be seen in Fig.~\ref{fig:Upsmmee},  an \PUps -signal is visible in both
\mpmm\ and \epem\ final states. The shape of the signals was derived from a
Monte Carlo simulation with
the relative production rates of the three \PUps -states fixed on
E605~\cite{E605} data. The background is described by a term for the
combinatorial background and a Drell-Yan term dominating at high
invariant masses.

As for the previous analyses, the cross section is determined relative to
the \PJgy\ cross section:
\begin{displaymath}
   \sigma_{\PUps\ } = \sigma(\PJgy\ ) \cdot 
   \frac{n(\PUps\ )}{n(\PJgy\ )} \cdot 
   \frac{Br(\PJgy\ \ra\ \dilepton\ )}{Br(\PUps\ \ra\ \dilepton\ )}
  \cdot \frac{\epsilon (\PJgy\ )}{\epsilon (\PUps )}
\end{displaymath}
The results of both subsamples are in good agreement. We obtain a combined
production cross section times branching ratio at mid-rapidity of 
$Br \cdot \frac{d\sigma}{dy} |_{y=0} = (3.4 \pm 0.8) $ pb/N.
In Fig.~\ref{fig:Upsfit} the \hb\  measurement is compared to the results of 
other experiments. The new point from \hb\  will help to clarify the
experimental situation which is controversial
in the energy regime of $\sqrt{s} \sim $ 40 GeV.
The energy dependence of the cross section can be described with the
parameterization~\cite{craigie}
\mbox{ 
$Br \cdot \frac{d\sigma}{dy} |_{y=0} = \sigma_0 \cdot \exp(-\frac{m_0}{\sqrt{s}}) $}.

\section{Summary}

The final data taking period of the \hb\ experiment was used to accumulate
150 million events with its di-lepton trigger. These events were produced
in pC and pW interactions of the 920 GeV proton beam of Hera.
The di-lepton sample includes
300,000 events with a \PJgy\ decaying into electon- or muon-pairs.
Studies of the dependence of the \PJgy\ production cross section on its
kinematic variables as well as on the target atomic number were shown
which turn to advantage the large acceptance in \mt\ and the backward
hemisphere.  
Preliminary results of the \PJzs\ to \PJgy\ as well as
\PChi\ to \PJgy\ production cross section ratios were shown.
New results on measurements of the \bbbar\ and \PUps\ production
cross section were also discussed.

\end{document}